\documentclass[journal=jctcce,manuscript=article]{achemso}

\usepackage[version=3]{mhchem} 
\usepackage{dcolumn}
\usepackage{multirow}
\usepackage{chngpage}
\newcolumntype{d}[1]{D{.}{.}{#1} }

\newcommand*{\citen}[1]{%
  \begingroup
    \romannumeral-`\x 
    \setcitestyle{numbers}%
    \cite{#1}%
  \endgroup   
}

\author{J. Matthias Kahk}
  \email{juhan.matthias.kahk@ut.ee}
 \affiliation{Institute of Physics, University of Tartu, W. Ostwaldi 1, 50411 Tartu, Estonia}
 
\author{Johannes Lischner}
\affiliation{Department of Physics and Department of Materials, and the Thomas Young Centre for Theory and Simulation of Materials, Imperial College London, London SW7 2AZ, United Kingdom}

\title[An \textsf{achemso} demo]
  {Combining the $\Delta$-Self-Consistent-Field and GW Methods for Predicting Core Electron Binding Energies in Periodic Solids}

\abbreviations{}

\begin{document}

\begin{tocentry}
	\includegraphics[width=8.255cm]{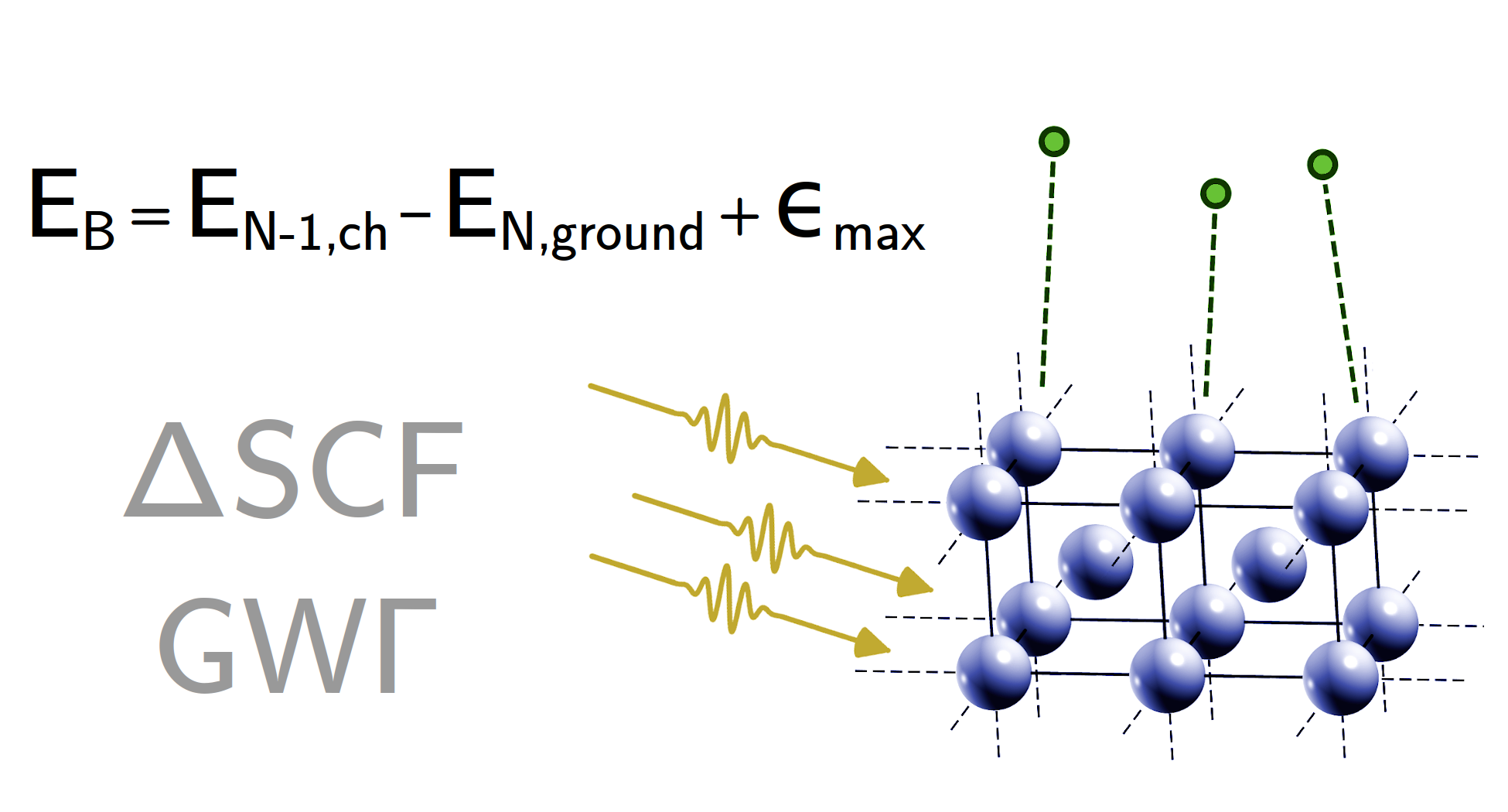}
\end{tocentry}

\begin{abstract}
  For the computational prediction of core electron binding energies in solids, two distinct kinds of modelling strategies have been pursued: the $\Delta$-Self-Consistent-Field method based on density functional theory (DFT), and the GW method. In this study, we examine the formal relationship between these two approaches, and establish a link between them. The link arises from the equivalence, in DFT, between the total energy difference result for the first ionization energy, and the eigenvalue of the highest occupied state, in the limit of infinite supercell size. This link allows us to introduce a new formalism, which highlights how in DFT - even if the total energy difference method is used to calculate core electron binding energies - the accuracy of the results still implicitly depends on the accuracy of the eigenvalue at the valence band maximum in insulators, or at the Fermi level in metals. We examine, whether incorporating a quasiparticle correction for this eigenvalue from GW theory improves the accuracy of the calculated core electron binding energies, and find that the inclusion of vertex corrections is required for achieving quantitative agreement with experiment.
\end{abstract}


\section{Introduction}

The energy required to remove a core electron from a particular atom depends on the atom’s chemical environment. In core level X-ray Photoelectron Spectroscopy (XPS), this dependence can be exploited to identify the chemical environments that are present in the sample. XPS is particularly well suited for the analysis of complex surfaces, and it plays an important role in the study of heterogeneous catalysis,\cite{salmeron_surfaces_2018,zhong_mini_2019,martin_catalytic_2022,degerman_operando_2022} corrosion,\cite{de_alwis_probing_2023,diler_initial_2014,hayez_xps_2004} environmental degradation,\cite{popescu_xps_2009,hahn_situ_2021,ries_xps_2019} or the manufacture of surface coatings \cite{kokkonen_ambient_2022,zanders_peald_2019,kisand_antimicrobial_2022,berens_effects_2020}. However, the interpretation of XPS spectra is challenging, which has motivated the development of computational techniques for calculating core electron binding energies from first principles \cite{bagus_self-consistent-field_1965,banna_study_1977,stener_lcao_1994,bagus_interpretation_2013,kahk_accurate_2019,hait_highly_2020,klein_nuts_2021,jana_slater_2022,zheng_performance_2019,arias-martinez_accurate_2022,golze_core-level_2018,golze_accurate_2020,mejia-rodriguez_scalable_2021,matthews_eom-cc_2020,hirao_vertical_2022,zhao_universal_2017}. 

For the prediction of absolute core electron binding energies in periodic solids, two kinds of methods have emerged. In the total energy difference method based on density functional theory (DFT), also known as the $\Delta$-Self-Consistent-Field ($\Delta$SCF) method, the core electron binding energy is calculated as the difference between total energies from two separate calculations: one for the system with a core hole, and one for the system without it \cite{kahk_core_2021,ozaki_absolute_2017,walter_offset-corrected_2016}. In contrast, in the GW method, the core electron binding energy is calculated as the GW eigenvalue of the relevant core eigenstate \cite{aoki_accurate_2018,zhu_all-electron_2021}. In a typical GW calculation, ground state orbitals and orbital eigenvalues are first obtained using DFT, and next, GW corrections to the eigenvalues are obtained by applying the GW method in a "one-shot" (G$_0$W$_0$), or partly self-consistent manner. Direct GW calculations of core electron binding energies involve some additional complications, when compared to GW calculations of valence states. Issues such as the treatment of the frequency-dependent self-energy, basis set convergence and extrapolation, starting point dependence, and the role of (partial) self-consistency have been discussed extensively in recent works \cite{golze_core-level_2018,golze_accurate_2020,li_benchmark_2022,mejia-rodriguez_scalable_2021,mejia-rodriguez_basis_2022}. In brief, very promising results have recently been obtained for molecular systems (mean absolute error $<$ 0.3 eV) [\citen{li_benchmark_2022}], whereas somewhat larger mean absolute errors (0.53 eV and 0.57 eV in references [\citen{aoki_accurate_2018}] and [\citen{zhu_all-electron_2021}], respectively) have been observed in the few preliminary studies of periodic solids published thus far.

In this work, we examine the formal relationship between the $\Delta$SCF and GW methods, and combine the two approaches by establishing the link between total energy differences and energy eigenvalues. In addition, we examine how this insight can be exploited to improve the accuracy of calculated binding energies.

\section{The $\Delta$SCF Method for Periodic Solids}

When calculating or measuring core electron binding energies in solids, a well-defined point of reference must be used. In experimental XPS, the sample Fermi level is typically used as the zero of the energy scale. However, as discussed in [\citen{kahk_core_2021}], this choice is not well suited for theoretical calculations of core electron binding energies in insulators, as the position of the Fermi level within the band gap is not in general known \textit{a priori}, and it depends strongly on extrinsic factors, such as the concentration of defects or impurities in the sample. Therefore, in recent computational studies, the energy of the highest occupied state, i.e. the Fermi level in metals and the valence band maximum (VBM) in insulators, has been used as the point of reference instead \cite{kahk_core_2021,aoki_accurate_2018,zhu_all-electron_2021}. 

For total energy difference methods, this means that the core electron binding energy is defined as the difference between two total energy differences: the $\Delta$SCF result for the core electron binding energy, and the $\Delta$SCF result for the first ionization energy of the solid. In the end, the total energy of the ground state cancels out:

\begin{equation}
E_\mathrm{B} = (E_{N-1,\mathrm{ch}} - E_{N,\mathrm{ground}}) - (E_{N-1,\mathrm{ground}} - E_{N,\mathrm{ground}}) = E_{N-1,\mathrm{ch}} - E_{N-1,\mathrm{ground}},
\label{eqn1}
\end{equation}

where $E_\mathrm{B}$ is the calculated core electron binding energy relative to the VBM in insulators or the Fermi level in metals, $E_{N,\mathrm{ground}}$ is the ground state total energy, $E_{N-1,\mathrm{ground}}$ is the total energy of the system with one electron removed from the highest occupied state, and $E_{N-1,\mathrm{ch}}$ is the total energy of the system with a core hole.

This formalism was used to calculate absolute core electron binding energies in solids in reference [\citen{kahk_core_2021}]. It was shown that core electron binding energies from periodic $\Delta$SCF calculations based on DFT with the SCAN functional \cite{sun_strongly_2015} were in good agreement with experimental values. In particular, the mean absolute error was just 0.24 eV for a small test set of 15 core electron binding energies. However, in some cases, significantly larger errors were observed, e.g. the C 1s binding energy in diamond was overestimated by 0.39 eV, and the Be 1s and O 1s binding energies in BeO were in error by 0.79 eV and 1.16 eV, respectively. In reference [\citen{kahk_core_2021}], it was speculated that these errors arise from the inability of DFT to accurately predict the position of the VBM in wide band gap insulators.

\subsection{The VBM Energy in Density Functional Theory}

In this study, we investigate this matter further. At first, from Equation \ref{eqn1}, it would seem that the VBM energy in fact never needs to be explicitly calculated for obtaining the core electron binding energy. However, the second term in the brackets before simplification does correspond to a total energy difference calculation of the VBM energy. The relationship between the term $(E_{N-1,\mathrm{ground}} – E_{N,\mathrm{ground}})$, and the VBM Kohn-Sham eigenvalue in DFT, $\epsilon_\mathrm{max}$ has been previously discussed, e.g. in [\citen{corsetti_system-size_2011}] and [\citen{persson_n_2005}]. In particular, as explained in [\citen{corsetti_system-size_2011}], the energy difference between a pure material and a material with a single hole becomes equal to the VBM Kohn-Sham eigenvalue in the limit of a dilute hole gas. In real calculations using finite supercells, however, this energy difference only slowly converges to the infinite limit as the system size is increased. Formally:

\begin{equation}
\lim\limits_{n \to \infty} (E_{N-1,\mathrm{ground}}(n) - E_{N,\mathrm{ground}}(n)) = -\epsilon_\mathrm{max},
\label{eqn2}
\end{equation}

where $n$ is the number of atoms per supercell, and $\epsilon_\mathrm{max}$ is the energy of the highest occupied state, i.e. VBM eigenvalue in insulators, or the eigenvalue at the Fermi level in metals. The preceding discussion pertains to DFT with real (approximate) exchange-correlation functionals. In exact DFT, the equality in Equation \ref{eqn2} holds at any supercell size. Equation \ref{eqn2} shows that in solids, at the limit of infinite supercell size, VBM energies calculated as total energy differences must have exactly the same shortcomings as Kohn-Sham eigenvalues.

\subsection{An Alternative Formalism for Periodic $\Delta$SCF Calculations of Core Electron Binding Energies}

Equation \ref{eqn2} allows us to write an alternative expression for the core electron binding energy, by replacing the term $(E_{N-1,\mathrm{ground}}-E_{N,\mathrm{ground}})$ with $-\epsilon_\mathrm{max}$:

\begin{equation}
E_\mathrm{B} = E_{N-1,\mathrm{ch}} - E_{N,\mathrm{ground}} +\epsilon_\mathrm{max}.
\label{eqn3}
\end{equation}

In the limit of infinite supercell size, Equation \ref{eqn1} and Equation \ref{eqn3} should give the same result, but for finite supercells the calculated core electron binding energies differ. A numerical verification of Equations \ref{eqn2} and \ref{eqn3} is presented next.

\subsection{Numerical Verification of Equation \ref{eqn2}}

\begin{figure}
	\centering
	\includegraphics[width=8.6cm]{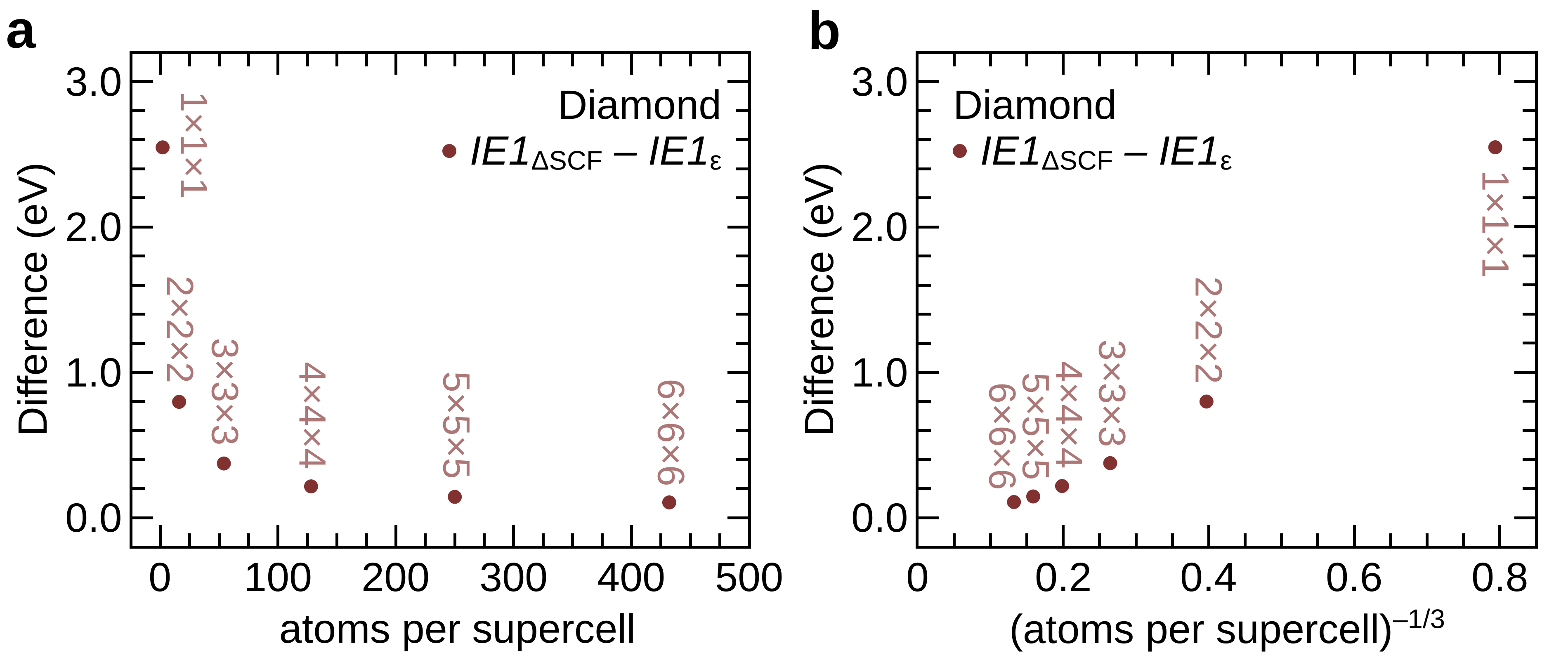}
	\caption{Numerical validation of Equation \ref{eqn2} for diamond. In panel (a), the difference between the first ionization energy calculated using the total energy difference method, and the negative eigenvalue of the highest occupied state is plotted against the number of atoms in the supercell. As the size of the supercell increases, the difference slowly tends towards zero. In panel (b), the same quantity is plotted against the inverse cube root of the number of atoms per supercell.}
	\label{Fig_DeltaSCF_vs_Eval}
\end{figure}

We have calculated $\mathit{IE}_{\mathrm{\Delta SCF}}$, defined as $E_{N-1,\mathrm{ground}}(n) - E_{N,\mathrm{ground}}(n)$ and $\mathit{IE}_{\epsilon}$, defined as $-\epsilon_\mathrm{max}$ for all of the 10 solids - Li, Be, Na, Mg, graphite, BeO, hex-BN, diamond, $\beta$-SiC, and Si - and all of the supercells considered in reference [\citen{kahk_core_2021}], using DFT with both the SCAN and the PBE functionals \cite{perdew_generalized_1996,sun_strongly_2015}. As an example, the results for diamond obtained using the SCAN functional are shown in Figure \ref{Fig_DeltaSCF_vs_Eval}. In Figure \ref{Fig_DeltaSCF_vs_Eval}(a), $\mathit{IE}_{\mathrm{\Delta SCF}} - \mathit{IE}_{\epsilon}$ is plotted against the number of atoms per supercell ($n$), and in Figure \ref{Fig_DeltaSCF_vs_Eval}(b), the same quantity is plotted against the inverse cube root of $n$, as is done when extrapolating core electron binding energies to the infinite supercell limit. Figures \ref{Fig_DeltaSCF_vs_Eval}(a) and \ref{Fig_DeltaSCF_vs_Eval}(b) show that $\mathit{IE}_{\mathrm{\Delta SCF}} - \mathit{IE}_{\epsilon}$ indeed slowly approaches zero as the size of the supercell increases. Similar behaviour is also observed for the other materials, using both PBE and SCAN – the detailed results are provided in the SI.

\subsection{Numerical Verification of Equation \ref{eqn3}}

\begin{figure}
	\centering
	\includegraphics[width=8.6cm]{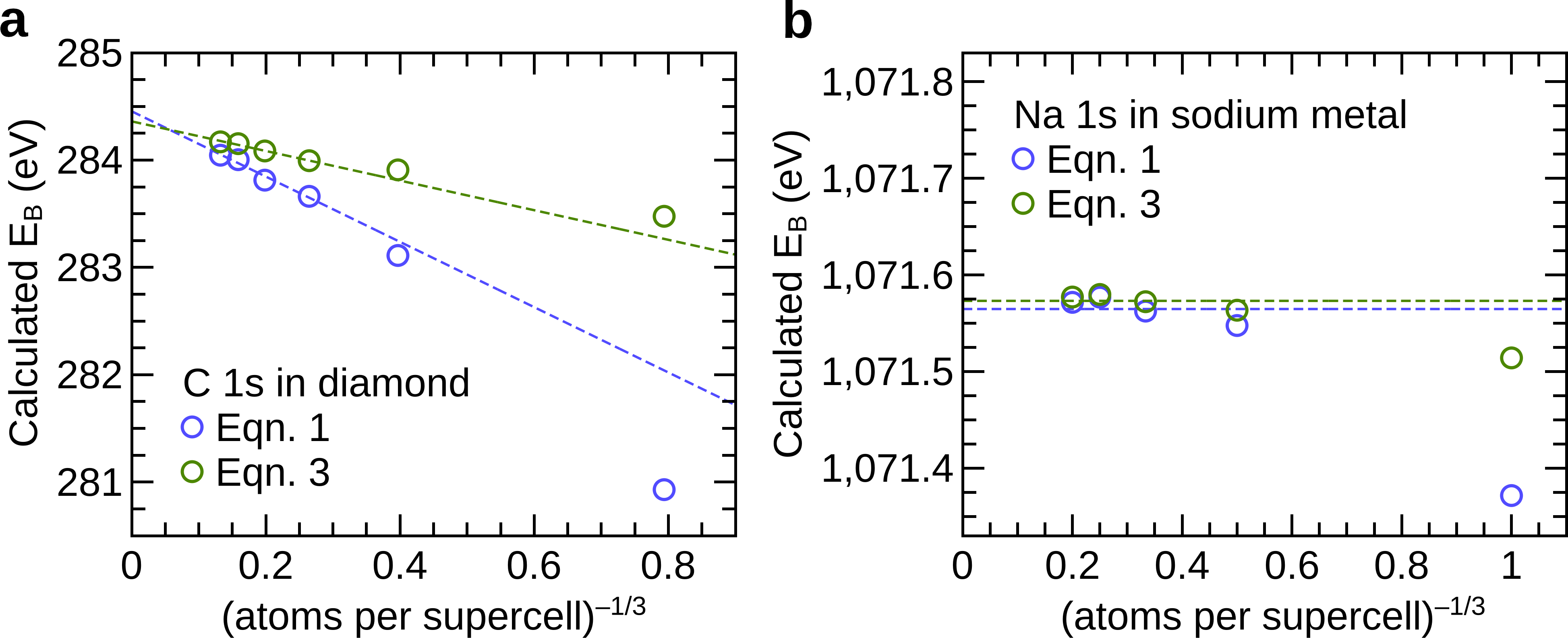}
	\caption{A comparison of calculated core electron binding energies from Equation \ref{eqn1} and Equation \ref{eqn3} for the C 1s level in diamond (panel a), and the Na 1s core level in sodium metal (panel b). For finite supercells, Equations \ref{eqn1} and \ref{eqn3} can give different results. However, at the limit of infinite supercell size, the calculated binding energies from Equation \ref{eqn1} and Equation \ref{eqn3} converge to the same limiting value.}
	\label{Fig_Eqn1_vs_Eqn3}
\end{figure}

Next, the core electron binding energies calculated using Equation \ref{eqn1} and Equation \ref{eqn3} are compared in Figure \ref{Fig_Eqn1_vs_Eqn3}. In Figure \ref{Fig_Eqn1_vs_Eqn3}, calculated core electron binding energies in one insulator, diamond, and one metal, Na, are shown as a function of supercell size. In each plot, the infinite supercell limit lies at the y-axis intercept. Figure \ref{Fig_Eqn1_vs_Eqn3}(a) shows calculated C 1s binding energies in diamond from Equation \ref{eqn1} and Equation \ref{eqn3}. The extrapolated values, 284.43 eV and 284.36 eV, respectively, differ by 0.07 eV - this is attributed to uncertainties in extrapolation and errors caused by finite k-point sampling. Whilst not negligible, this difference is less than half of the average error in the calculated binding energies, and of the same magnitude as the precision with which experimental binding energies are typically reported. In Figure \ref{Fig_Eqn1_vs_Eqn3}(b), the calculated Na 1s binding energies in Na metal from the two equations are compared. In this case, and similarly for other metals, for sufficiently large supercells, both equations yield core electron binding energies that are converged to the limiting value. For the Na 1s binding energy in Na metal, the limiting values from Equation \ref{eqn1} and Equation \ref{eqn3} differ by less than 0.01 eV.

Further numerical verification of Equation \ref{eqn3} is provided in Table \ref{Table_Eqn1_vs_Eqn2}, where a comparison of the extrapolated results from Equation \ref{eqn1} and Equation \ref{eqn3} is provided for all of the 15 core levels considered in ref. [\citen{kahk_core_2021}]. The same calculations have been performed using both the PBE and SCAN exchange-correlation functionals. In summary, the two equations yield very similar results. For both SCAN and PBE, the root mean squared deviation between the calculated binding energies from the two equations is just 0.07 eV.

\setlength{\tabcolsep}{10pt}
\begin{table}
    \caption{A comparison of core electron binding energies, extrapolated to the infinite supercell limit, from Equation \ref{eqn1} and Equation \ref{eqn3}. The results are shown for two sets of calculations, one using the exchange-correlation functional PBE, and the other using the exchange-correlation functional SCAN. All energies are given in eV.}
    \begin{tabular}{ c c d{1} d{1} d{1} d{1} d{1} d{1} }
        \multirow{2}{*}{Solid} & \multirow{2}{*}{Core level} & \multicolumn{3}{c}{$E\mathrm{_B}$ (PBE)} & \multicolumn{3}{c}{$E\mathrm{_B}$ (SCAN)} \\
        & & \multicolumn{1}{c}{Eqn. 1} & \multicolumn{1}{c}{Eqn. 3} & \multicolumn{1}{c}{diff.} & \multicolumn{1}{c}{Eqn. 1} & \multicolumn{1}{c}{Eqn. 3} & \multicolumn{1}{c}{diff.} \\
        \hline
        Li & Li 1s & 54.64 & 54.64 & 0.00 & 54.88 & 54.87 & 0.01 \\
        Be & Be 1s & 111.43 & 111.48 & -0.05 & 111.88 & 111.91 & -0.03\\
        Na & Na 1s & 1,069.67 & 1,069.68 & -0.01 & 1,071.56 & 1,071.59 & -0.03\\
        Na & Na 2p & 30.57 & 30.58 & -0.01 & 30.65 & 30.66 & -0.01\\
        Mg & Mg 1s & 1,300.88 & 1,300.89 & -0.01 & 1,303.25 & 1,303.26 & -0.01\\
        Mg & Mg 2p & 49.44 & 49.44 & 0.00 & 49.69 & 49.74 & -0.05\\
        Graphite & C 1s & 283.63 & 283.44 & 0.19 & 284.44 & 284.19 & 0.25\\
        BeO & Be 1s & 110.45 & 110.44 & 0.01 & 110.79 & 110.78 & 0.01\\
        BeO & O 1s & 528.20 & 528.18 & 0.02 & 528.86 & 528.83 & 0.03\\
        hex-BN & B 1s & 187.73 & 187.73 & 0.00 & 188.42 & 188.44 & -0.02\\
        hex-BN & N 1s & 395.75 & 395.71 & 0.04 & 396.39 & 396.36 & 0.03\\
        Diamond & C 1s & 283.97 & 283.80 & 0.17 & 284.43 & 284.36 & 0.07\\
        beta-SiC & Si 2p & 98.76 & 98.72 & 0.04 & 99.24 & 99.19 & 0.05\\
        beta-SiC & C 1s & 280.93 & 280.92 & 0.01 & 281.48 & 281.44 & 0.04\\
        Si & Si 2p & 98.73 & 98.64 & 0.09 & 99.17 & 99.17 & 0.00\\
        \hline
        \multicolumn{4}{r}{Maximum:} & 0.19 & & & 0.25 \\
        \multicolumn{4}{r}{Mean:} & 0.03 & & & 0.02 \\
        \multicolumn{4}{r}{Root mean squared:} & 0.07 & & & 0.07 \\
        \hline
    \end{tabular}
    \label{Table_Eqn1_vs_Eqn2}
\end{table}

\subsection{Localized vs. Delocalized Hole States}

\begin{figure}
	\centering
	\includegraphics[width=8.6cm]{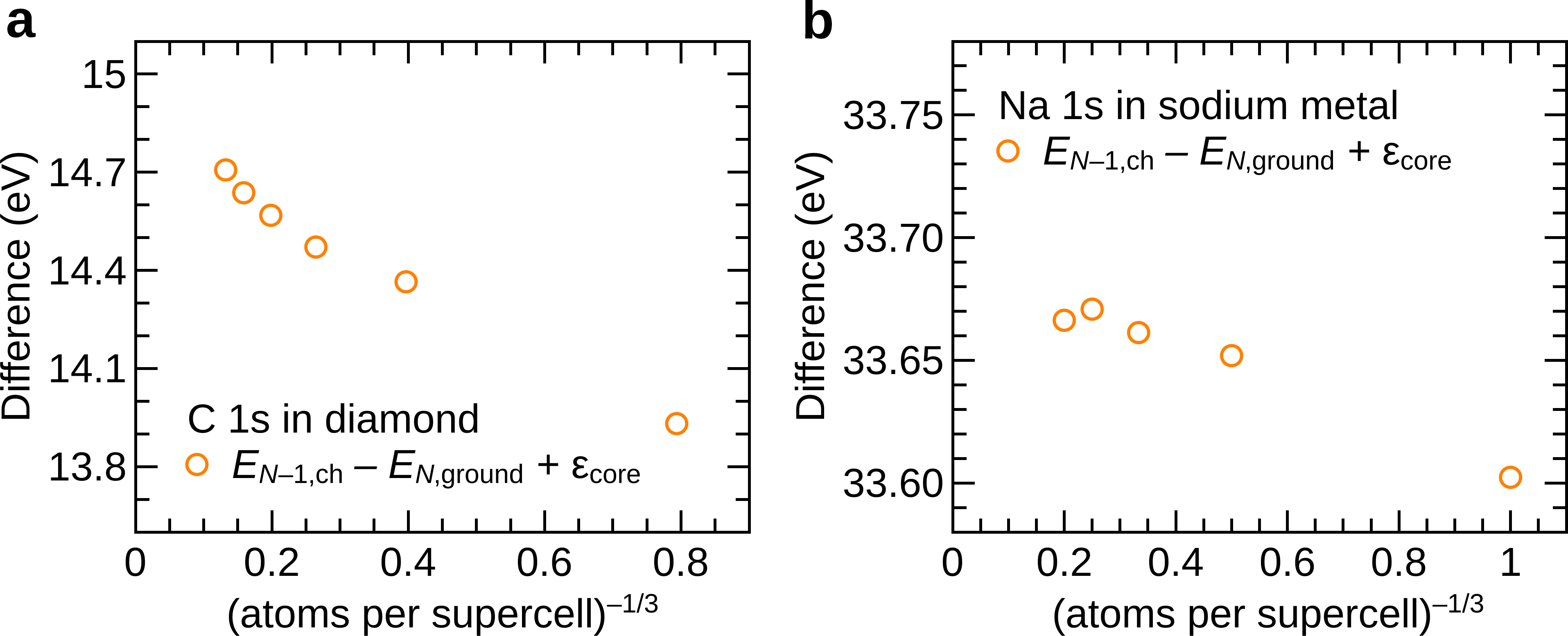}
	\caption{The difference between the calculated core electron binding energy from a total energy difference calculation, and the negative eigenvalue of the core orbital, as a function of supercell size. Results for the C 1s core level in diamond are shown in panel a, and results for the Na 1s core level in sodium metal are shown in panel b. In contrast to the behaviour observed for the first ionization energy (Figure \ref{Fig_DeltaSCF_vs_Eval}), for core electron binding energies, the difference does not approach zero with increasing supercell size. This is due to the localized nature of the core hole, as opposed to the delocalized nature of the hole in the valence band.}
	\label{Fig_core_DeltaSCF_vs_E_val}
\end{figure}

It is important to emphasize that an identity similar to Equation \ref{eqn2} does not hold for the core electrons, i.e. 

\begin{equation}
\lim\limits_{n \to \infty} (E_{N-1,\mathrm{ch}}(n) - E_{N,\mathrm{ground}}(n)) \neq -\epsilon_{\mathrm{core}},
\label{eqn4}
\end{equation}

provided that the core hole is properly localized in the calculation of $E_{N-1,\mathrm{ch}}$. This is numerically illustrated in Figure \ref{Fig_core_DeltaSCF_vs_E_val}. This fundamental difference arises due to the fact that in valence ionization an electron is removed from a delocalized state, and as the size of the simulation cell increases, the change in the local potential experienced by all the remaining electrons slowly tends towards zero. In contrast, in core ionization, an electron is removed from a localized state, and in the vicinity of the atom with a core hole, the remaining electrons experience a large change in local potential regardless of the size of the supercell. Here, the terms "localized" and "delocalized" refer to the spatial distribution of a Kohn-Sham state relative to the simulation cell (that in general contains many unit cells of the solid). A localized core hole is centred around exactly one atom, regardless the size of the simulation cell, thus breaking the translational symmetry in the system. In contrast, a hole in a delocalized state is evenly distributed over all symmetry-equivalent atoms in the simulation cell.

\section{The Significance of Equation 3}

Conceptually, Equation \ref{eqn3} highlights that the accuracy of core electron binding energies from periodic $\Delta$SCF calculations depends on the accuracy of $\epsilon_\mathrm{max}$, i.e. the DFT eigenvalue of the highest occupied state. However, DFT is widely known to underestimate band gaps in solids, and more advanced theories such as the GW approximation yield significant corrections to both the VBM and CBM (conduction band minimum) energies predicted by DFT. It is therefore reasonable to consider, whether it is possible to improve the accuracy of calculated core electron binding energies in insulating solids by adding a quasiparticle correction to $\epsilon_\mathrm{max}$ in Equation \ref{eqn3}. 

In other words, provided that a consistent point of reference can be established, one could try to calculate $(E_{N-1,\mathrm{ch}} - E_{N,\mathrm{ground}})$ using a method that is optimal for predicting core electron binding energies, and $\epsilon_{\mathrm{max}}$ using a method that is optimal for modelling the removal of valence electrons, and combine the two values to obtain a “theoretical best estimate” core electron binding energy referenced to the VBM (or $E_\mathrm{F}$ in metals).

\section{Combining the $\Delta$SCF and GW approaches}

In this work, we attempt to combine core electron binding energies calculated using the $\Delta$SCF method with VBM energies calculated using the G$_0$W$_0$ approach. In particular, we have performed the following calculations.

(i) We have calculated $(E_{N-1,\mathrm{ch}} - E_{N,\mathrm{ground}})$, as well as $\epsilon_{\mathrm{max}}^{\mathrm{DFT}}$, for all of the materials, core levels, and supercells considered in [\citen{kahk_core_2021}], using DFT with two different functionals: PBE and SCAN. These calculations have been performed in the all-electron electronic structure code FHI-aims \cite{blum_ab_2009}. Further details are provided in methods section.

(ii) We have calculated $\epsilon_{\mathrm{max}}^{\mathrm{PBE}}$ and $\epsilon_{\mathrm{max}}^{\mathrm{G_{0}W_{0}@PBE}}$ for each of the solids using the electronic structure code GPAW. Details of these calculations are also provided in the methods section. The G$_0$W$_0$ correction to the eigenvalue of the highest occupied state, $\Delta E_{\mathrm{G_{0}W_{0}@PBE}}$, is defined as $\epsilon_{\mathrm{max}}^{\mathrm{G_{0}W_{0}@PBE}} - \epsilon_{\mathrm{max}}^{\mathrm{PBE}}$.

(iii) Combining the G$_0$W$_0$ correction with $\Delta$SCF core electron binding energies calculated with PBE is straightforward. The corrected binding energy is obtained as $E_B = E_{N-1,\mathrm{ch}}^{\mathrm{PBE}} – E_{N,\mathrm{ground}}^{\mathrm{PBE}} + \epsilon_\mathrm{max}^\mathrm{PBE} + \Delta E_{\mathrm{G_{0}W_{0}@PBE}}$.

The total energies $E_{N-1,\mathrm{ch}}$ and $E_{N,\mathrm{ground}}$, as well as $\epsilon_\mathrm{max}$ have all been calculated in FHI-aims, using the same structures, physical settings (functional and treatment of relativistic effects), and numerical settings (basis sets, integration grids, etc.).

(iv) We have also attempted to combine a G$_0$W$_0$ correction with the core electron binding energies calculated using the SCAN functional. For technical reasons, and due to the limited current knowledge about the performance of DFT with the SCAN functional as a starting point for perturbative GW calculations, we have not at present calculated G$_0$W$_0$ corrections to the VBM (or Fermi level) eigenvalues from SCAN. Instead, we have chosen to test a strategy where the G$_0$W$_{0}@$PBE correction is combined with core electron binding energies from $\Delta$SCF calculated using the SCAN functional. This requires an additional step, because the correction is defined relative to the PBE eigenvalue of the highest occupied state, not the SCAN eigenvalue. Therefore, we also have to correct for the difference between $\epsilon_\mathrm{max}^\mathrm{PBE}$ and $\epsilon_\mathrm{max}^\mathrm{SCAN}$, and the corrected binding energies are obtained as $E_B = E_{N-1,\mathrm{ch}}^{\mathrm{SCAN}} - E_{N,\mathrm{ground}}^{\mathrm{SCAN}} + \epsilon_{\mathrm{max}}^{\mathrm{SCAN}} + \Delta E_\mathrm{PBE@SCAN} + \Delta E_{\mathrm{G_{0}W_{0}@PBE}}$.

Here, $\Delta E_{\mathrm{PBE@SCAN}}$ refers to $\epsilon_\mathrm{max}^\mathrm{PBE@SCAN}$ - $\epsilon_\mathrm{max}^\mathrm{SCAN}$, where $\epsilon_\mathrm{max}^\mathrm{PBE@SCAN}$ is the VBM eigenvalue from PBE evaluated non-self-consistently using the Kohn-Sham orbitals from a converged ground state calculation with the SCAN functional. There is a conceptual difficulty with this approach, namely that $\epsilon_{\mathrm{max}}^{\mathrm{SCAN}}$ and $\Delta E_{\mathrm{PBE@SCAN}}$ are evaluated at the optimized density from SCAN, whereas $\Delta E_{\mathrm{G_{0}W_{0}@PBE}}$ is evaluated at the optimized density from PBE. In order to assess the severity of this approximation, we have compared $\Delta E_\mathrm{PBE@SCAN}$ with $\Delta E_\mathrm{SCAN@PBE}$, for each of the materials considered, i.e. the differences between the SCAN and PBE eigenvalues at the relaxed density from either functional. We have found that $\Delta E_{\mathrm{PBE@SCAN}} \approx -\Delta E_{\mathrm{SCAN@PBE}}$ in all cases, with all differences in the absolute values being less than 0.02 eV.

Thus we obtain (i) uncorrected core electron binding energies from Equation \ref{eqn3} using PBE and SCAN: $E_{\mathrm{B}}^{\mathrm{PBE}}$ and $E_{\mathrm{B}}^{\mathrm{SCAN}}$, and (iii,iv) core electron binding energies that have been recalibrated to the position of the highest occupied state predicted by the G$_0$W$_{0}@$PBE method: $E_{\mathrm{B}}^{\mathrm{PBE,\Delta \mathit{E}_{G_{0}W_{0}@PBE}}}$, and $E_{\mathrm{B}}^{\mathrm{SCAN,\Delta \mathit{E}_{G_{0}W_{0}@PBE}}}$. The initial results obtained using this approach are disappointing. In fact, as shown in Tables \ref{Table_PBE_BE_and_corr} and \ref{Table_SCAN_BE_and_corr}, including the correction for $\epsilon_\mathrm{max}$ from G$_0$W$_0$ theory worsens the agreement with experiment considerably.

\setlength{\tabcolsep}{6pt}
\begin{table}
    \begin{adjustwidth}{-.8in}{-.8in}
    \begin{center}
    \caption{Core electron binding energies from $\Delta$SCF calculations based on Equation \ref{eqn3} and the PBE functional, and from calculations where a G$_0$W$_0$ or G$_0$W$_{0}\Gamma$ correction has been applied to $\epsilon_\mathrm{max}$ in Equation \ref{eqn3}. All energies are given in eV.}
    \small
    \begin{tabular}{ c c d{1} d{1} d{1} d{1} d{1} d{1} d{1}}
        \multirow{2}{*}{Solid} & \multicolumn{1}{c}{Core} & \multicolumn{1}{c}{$E_\mathrm{B}$} & \multirow{2}{*}{$E_{\mathrm{B}}^{\mathrm{PBE}}$} & \multirow{2}{*}{Error} & \multirow{2}{*}{$E_{\mathrm{B}}^{\mathrm{PBE,\Delta \mathit{E}_{G_{0}W_{0}@PBE}}}$} & \multirow{2}{*}{Error} & \multirow{2}{*}{$E_{\mathrm{B}}^{\mathrm{PBE,\Delta \mathit{E}_{G_{0}W_{0}\Gamma@PBE}}}$} & \multirow{2}{*}{Error} \\
        & \multicolumn{1}{c}{level} & \multicolumn{1}{c}{Expt.} & & & & & & \\
        \hline
        Li & Li 1s & 54.85 & 54.64 & -0.21 & 54.54 & -0.31 & 54.71 & -0.14 \\
        Be & Be 1s & 111.85 & 111.48 & -0.37 & 111.21 & -0.64 & 111.97 & 0.12 \\
        Na & Na 1s & 1071.75 & 1069.68 & -2.07 & 1069.37 & -2.38 & 1069.79 & -1.96 \\
        Na & Na 2p & 30.51 & 30.58 & 0.07 & 30.27 & -0.24 & 30.69 & 0.18 \\
        Mg & Mg 1s & 1303.24 & 1300.89 & -2.35 & 1300.44 & -2.80 & 1301.10 & -2.14 \\
        Mg & Mg 2p & 49.79 & 49.44 & -0.35 & 48.99 & -0.80 & 49.65 & -0.14 \\
        Graphite & C 1s & 284.41 & 283.44 & -0.97 & 283.02 & -1.39 & 283.77 & -0.64 \\
        BeO & Be 1s & 110.00 & 110.44 & 0.44 & 108.17 & -1.83 & 108.56 & -1.44 \\
        BeO & O 1s & 527.70 & 528.18 & 0.48 & 525.91 & -1.79 & 526.30 & -1.40 \\
        hex-BN & B 1s & 188.35 & 187.73 & -0.62 & 186.29 & -2.06 & 186.89 & -1.46 \\
        hex-BN & N 1s & 396.00 & 395.71 & -0.29 & 394.27 & -1.73 & 394.87 & -1.13 \\
        Diamond	& C 1s & 284.04 & 283.80 & -0.24  & 282.57 & -1.47 & 283.35 & -0.69 \\
        beta-SiC & Si 2p & 99.20 & 98.72 & -0.48 & 97.68 & -1.52 & 98.40 & -0.80 \\
        beta-SiC & C 1s & 281.55 & 280.92 & -0.63 & 279.88 & -1.67 & 280.60 & -0.95 \\
        Si & Si 2p & 99.03 & 98.64 & -0.39 & 97.95 & -1.08 & 98.65 & -0.38 \\
        \hline
        \multicolumn{4}{r}{Mean error:} & -0.53 & & -1.45 & & -0.86 \\
        \multicolumn{4}{r}{Mean absolute error:} & 0.66 & & 1.45 & & 0.90 \\
        \hline
    \end{tabular}
    \label{Table_PBE_BE_and_corr}
    \end{center}
    \end{adjustwidth}
\end{table}

\begin{table}
    \begin{adjustwidth}{-.8in}{-.8in}
    \begin{center}
    \caption{Core electron binding energies from $\Delta$SCF calculations based on Equation \ref{eqn3} and the SCAN functional, and from calculations where a G$_0$W$_0$ or G$_0$W$_{0}\Gamma$ correction has been applied to $\epsilon_\mathrm{max}$ in Equation \ref{eqn3}. In this case the correction consists of two parts: $\Delta E_\mathrm{PBE@SCAN}$ shifts a binding energy onto a scale where the zero is defined by the position of the VBM predicted by PBE, and $\Delta E_{\mathrm{G_{0}W_{0}@PBE}}$ ($\Delta E_{\mathrm{G_{0}W_{0}\Gamma@PBE}}$) shifts it further onto a scale where the zero is defined by the position of the VBM predicted by G$_0$W$_0$@PBE (G$_0$W$_{0}\Gamma$@PBE). All energies are given in eV.}
    \small
    \begin{tabular}{ c c d{1} d{1} d{1} d{1} d{1} d{1} d{1}}
        \multirow{2}{*}{Solid} & \multicolumn{1}{c}{Core} & \multicolumn{1}{c}{$E_\mathrm{B}$} & \multirow{2}{*}{$E_{\mathrm{B}}^{\mathrm{SCAN}}$} & \multirow{2}{*}{Error} & \multirow{2}{*}{$E_{\mathrm{B}}^{\mathrm{SCAN,\Delta \mathit{E}_{G_{0}W_{0}@PBE}}}$} & \multirow{2}{*}{Error} & \multirow{2}{*}{$E_{\mathrm{B}}^{\mathrm{SCAN,\Delta \mathit{E}_{G_{0}W_{0}\Gamma@PBE}}}$} & \multirow{2}{*}{Error} \\
        & \multicolumn{1}{c}{level} & \multicolumn{1}{c}{Expt.} & & & & & & \\
        \hline
        Li & Li 1s & 54.85 & 54.87 & 0.02 & 54.68 & -0.17 & 54.85 & 0.00 \\
        Be & Be 1s & 111.85 & 111.91 & 0.06 & 111.58 & -0.27 & 112.34 & 0.49 \\
        Na & Na 1s & 1071.75 & 1071.59 & -0.16 & 1071.28 & -0.47 & 1071.70 & -0.05 \\
        Na & Na 2p & 30.51 & 30.66 & 0.15 & 30.35 & -0.16 & 30.77 & 0.26 \\
        Mg & Mg 1s & 1303.24 & 1303.26 & 0.02 & 1302.82 & -0.42 & 1303.48 & 0.24 \\
        Mg & Mg 2p & 49.79 & 49.74 & -0.05 & 49.30 & -0.49 & 49.96 & 0.17 \\
        Graphite & C 1s & 284.41 & 284.19 & -0.22 & 283.77 & -0.64 & 284.53 & 0.12 \\
        BeO & Be 1s & 110.00 & 110.78 & 0.78 & 109.15 & -0.85 & 109.54 & -0.46 \\
        BeO & O 1s & 527.70 & 528.83 & 1.13 & 527.20 & -0.50 & 527.59 & -0.11 \\
        hex-BN & B 1s & 188.35 & 188.44 & 0.09 & 187.41 & -0.94 & 188.02 & -0.33 \\
        hex-BN & N 1s & 396.00 & 396.36 & 0.36 & 395.33 & -0.67 & 395.94 & -0.06 \\
        Diamond & C 1s & 284.04 & 284.36 & 0.32 & 283.30 & -0.74 & 284.08 & 0.04 \\
        beta-SiC & Si 2p & 99.20 & 99.19 & -0.01 & 98.39 & -0.81 & 99.11 & -0.09 \\
        beta-SiC & C 1s & 281.55 & 281.44 & -0.11 & 280.64 & -0.91 & 281.36 & -0.19 \\
        Si & Si 2p & 99.03 & 99.17 & 0.14 & 98.65 & -0.38 & 99.36 & 0.33 \\
        \hline
        \multicolumn{4}{r}{Mean error:} & 0.17 & & -0.56 & & 0.02 \\
        \multicolumn{4}{r}{Mean absolute error:} & 0.24 & & 0.56 & & 0.19 \\
        \hline
    \end{tabular}
    \label{Table_SCAN_BE_and_corr}
    \end{center}
    \end{adjustwidth}
\end{table}

For PBE, the mean absolute error (MAE) increases from 0.66 eV to 1.45 eV, and for SCAN, the MAE increases from 0.24 eV to 0.56 eV. In particular, we find that if the G$_0$W$_0$ correction for the highest occupied state is included, the calculated binding energies are too low, as compared to experiment, in all cases. This means that the mean signed errors (MSE) are equal in magnitude to the mean abolute errors: -1.45 eV for $E_{\mathrm{B}}^{\mathrm{PBE,\Delta \mathit{E}_{G_{0}W_{0}@PBE}}}$, and -0.56 eV for $E_{\mathrm{B}}^{\mathrm{SCAN,\Delta \mathit{E}_{G_{0}W_{0}@PBE}}}$. In contrast, the mean signed errors for $E_\mathrm{B}^\mathrm{PBE}$ and $E_\mathrm{B}^\mathrm{SCAN}$ are a lot smaller: -0.53 eV and +0.17 eV respectively.

\subsection{The Effect of Vertex Corrections in GW}

In reference [\citen{schmidt_simple_2017}], it was argued that whilst the G$_0$W$_0$ method is highly accurate for band gaps in periodic solids, it relies partly on error cancellation, and that the absolute band energies predicted by G$_0$W$_0$ are considerably less accurate. An improved methodology, termed G$_0$W$_{0}\Gamma$, was proposed, in which so-called vertex corrections derived from the renormalized adiabatic local density approximation (rALDA) kernel are included. It was shown, that as compared to G$_0$W$_0$, the band gaps predicted by G$_0$W$_{0}\Gamma$ are largely unchanged, whereas the absolute positions of the band edges are shifted upwards by approximately 0.6 eV in the examples considered.

We have examined whether using the G$_0$W$_{0}\Gamma@$PBE correction to the energy of the highest occupied state, instead of the G$_0$W$_{0}@$PBE correction, improves the results. The respective binding energies are labelled $E_{\mathrm{B}}^{\mathrm{PBE,\Delta \mathit{E}_{G_{0}W_{0}\Gamma@PBE}}}$ and $E_{\mathrm{B}}^{\mathrm{SCAN,\Delta \mathit{E}_{G_{0}W_{0}\Gamma@PBE}}}$. The results shown in Tables \ref{Table_PBE_BE_and_corr} and \ref{Table_SCAN_BE_and_corr} indicate that the G$_0$W$_{0}\Gamma$ correction performs considerably better than the simpler G$_0$W$_0$ correction. For PBE, the corrected binding energies are still somewhat less accurate with the uncorrected results, with MAE = 0.90 eV. In contrast, the MAE for the $E_{\mathrm{B}}^{\mathrm{SCAN,\Delta \mathit{E}_{G_{0}W_{0}\Gamma@PBE}}}$ results is just 0.19 eV, which is smaller than the MAE of the uncorrected binding energies. Overall, the G$_0$W$_{0}\Gamma$ correction improves the accuracy of the calculated binding energies in non-metals: the MAE of the corrected binding energies is 0.19 eV, as compared to 0.35 eV for the pure $\Delta$SCF results with SCAN. In particular, the G$_0$W$_{0}\Gamma$ correction significantly improves the results for the difficult cases of diamond and BeO – the errors in the C 1s, Be 1s and O 1s binding energies are reduced to 0.04 eV, -0.46 eV, and -0.11 eV respectively, compared to 0.32 eV, 0.78 eV, and 1.13 eV for the $E_\mathrm{B}^\mathrm{SCAN}$ values. In the metallic systems considered in this work, the accuracy of the original $\Delta$SCF results with SCAN is already very high: MAE = 0.08 eV; the G$_0$W$_{0}\Gamma$ correction makes the agreement somewhat worse, although the MAE remains relatively small at 0.20 eV.

\section{Conclusions}

In summary, this study establishes a direct link between the two fundamentally different strategies that can be employed for calculating core electron binding energies: total energy difference methods, and eigenvalue methods. Formally, this is expressed as the equivalence of Equations \ref{eqn1} and \ref{eqn3} in the limit of infinite supercell size. The results indicate that combining a technique that is known to yield accurate absolute core electron binding energies in free molecules ($\Delta$SCF with SCAN) with an approach that yields accurate band energies of valence states (G$_0$W$_{0}\Gamma$) is a viable strategy for calculating core electron binding energies in solids, referenced to the energy of the highest occupied state. Nevertheless, the smallness of the dataset (only 15 binding energies) means that additional and more extensive tests are required to properly evaluate the accuracy of the SCAN + $\Delta E_\mathrm{G_{0}W_{0}\Gamma@PBE}$ approach. 

In more general terms, we have demonstrated the importance of accurately predicting the position of the VBM in calculations of core electron binding energies, whenever the VBM is used as a point of reference. This includes not only calculations of periodic solids, but also calculations of surface species adsorbed onto a substrate with a band gap. We have found that using the conventional G$_0$W$_0$ approach to predict the VBM energy gives unsatisfactory results. In contrast, using VBM energies predicted by the G$_0$W$_{0}\Gamma$ approach, in which vertex corrections are included, yields excellent agreement between the calculated and experimental core electron binding energies. Other strategies for going beyond the G$_0$W$_0$@PBE level of theory, such as using a different mean-field starting point, or including partial self-consistency in GW, may give similar improvements \cite{caruso_benchmark_2016,li_benchmark_2022}, and will be investigated in future studies. As an alternative with lower computational cost, hybrid functionals could be used to predict the VBM energy. This could be useful in cases where performing a GW calculation of the full unit cell of the material is prohibitively expensive.

\section{Computational Methods}

All of the $\Delta$SCF calculations were performed using the all-electron electronic structure code FHI-aims \cite{blum_ab_2009,yu_elsi_2018,havu_efficient_2009}. The results of the calculations reported in [\citen{kahk_core_2021}], based on the SCAN functional, have been reused in this work to calculate core electron binding energies based on Equation \ref{eqn1} and Equation \ref{eqn3}. In addition, similar calculations, using the same structures, settings, and numerical parameters have been run using the PBE functional. Full details are provided in the supplementary information of [\citen{kahk_core_2021}].

GW and GW$\Gamma$ calculations were run using GPAW \cite{mortensen_real-space_2005,enkovaara_electronic_2010,huser_quasiparticle_2013}. In these calculations, the valence electrons are modelled using a plane wave basis set, and the effect of core electrons is treated using the projector-augmented wave formalism, as described in [\citen{mortensen_real-space_2005,enkovaara_electronic_2010}]. In the ground state DFT calculations in GPAW, a plane wave cutoff of 800 eV was employed. The structures and the k-point grids used are given in the supplementary information. Occupation smearing based on the Fermi-Dirac distribution with a width of 0.001 eV was applied in all cases. In the GW and GW$\Gamma$ calculations, a non-linear frequency grid defined by the values $\omega_2 = 20$ eV and $\Delta\omega_0 = 0.02$ eV was used, where $\Delta\omega_0$ is the frequency spacing at $\omega = 0$ and $\omega_2$ is the frequency at which the spacing has increased to 2$\Delta\omega_0$. For GW$\Gamma$, vertex corrections were calculated using the rAPBE kernel. GW and GW$\Gamma$ calculations were performed at three values of $E_\mathrm{cut}$: 300 eV, 350 eV, and 400 eV, where $E_\mathrm{cut}$ is the plane wave cutoff, and converged values were obtained by using a $1/E_{\mathrm{cut}}^{3/2}$ extrapolation.

\begin{acknowledgement}

This project has received funding from the European Union's Horizon 2020 research and innovation programme under grant agreement No 892943. JMK acknowledges support from the Estonian Centre of Excellence in Research project “Advanced materials and high-technology devices for sustainable energetics, sensorics and nanoelectronics” TK141 (2014-2020.4.01.15-0011). This work used the ARCHER2 UK National Supercomputing Service via J.L.’s membership of the HEC Materials Chemistry Consortium of UK, which is funded by EPSRC (EP/L000202).

\end{acknowledgement}

\begin{suppinfo}

Binding energies from Equation 1 and Equation 3, and extrapolation plots, numerical verification of Equation 2 for all materials considered in this work, structures and k-point grids used in the GW and GW$\Gamma$ calculations, extrapolation of the GW and GW$\Gamma$ results to $E_\mathrm{cut} \xrightarrow{} +\infty$.

\end{suppinfo}

\bibliography{2nd_Periodic_BE_Paper_resubmission}

\end{document}